\documentclass[%
 aip,
 amsmath,amssymb,
 reprint,%
]{revtex4-1}

\usepackage{graphicx}
\usepackage{dcolumn}
\usepackage{bm}

\usepackage[utf8]{inputenc}
\usepackage[T1]{fontenc}
\usepackage{mathptmx}
\usepackage{etoolbox}
\newcommand{\BiSb}{$\mathrm{Bi_{1-x}Sb_{x}}$}
\bibpunct{[}{]}{,}{n}{,}{,}
\makeatletter
\def\@email#1#2{%
 \endgroup
 \patchcmd{\titleblock@produce}
  {\frontmatter@RRAPformat}
  {\frontmatter@RRAPformat{\produce@RRAP{*#1\href{mailto:#2}{#2}}}\frontmatter@RRAPformat}
  {}{}
}%
\makeatother
\begin{document}

\preprint{AIP/123-QED}

\title{Epitaxial growth and characterization of Bi$_{1-x}$Sb$_x$ thin films on (0001) sapphire substrates}
\author{Yu-Sheng Huang}
\author{Saurav Islam}\email{ski5160@psu.edu}
\author{Yongxi Ou}
 \affiliation{Department of Physics, Pennsylvania State University, University Park, Pennsylvania 16802, USA}
 \author{Supriya Ghosh}
\affiliation{%
Department of Chemical Engineering and Materials Science, University of Minnesota, Minneapolis, Minnesota 55455, USA
}%
\author{Anthony Richardella}
 \affiliation{Department of Physics and Materials Research Institute, Pennsylvania State University, University Park, Pennsylvania 16802, USA}
\author{K. Andre Mkhoyan}
\affiliation{%
Department of Chemical Engineering and Materials Science, University of Minnesota, Minneapolis, Minnesota 55455, USA
}%
\author{Nitin Samarth}
\email{nsamarth@psu.edu}
\affiliation{Department of Physics, Department of Materials Science and Engineering, and Materials Research Institute, Pennsylvania State University, University Park, Pennsylvania 16802, USA}

\date{\today}

\begin{abstract}

We report the molecular beam epitaxy of \BiSb~thin films ($0 \leq x \leq 1$) on sapphire (0001) substrates using a thin (Bi,Sb)$_2$Te$_3$ buffer layer. Characterization of the films using reflection high energy diffraction, x-ray diffraction, atomic force microscopy, and scanning transmission electron microscopy reveals epitaxial growth of films of reasonable structural quality. This is further confirmed via x-ray diffraction pole figures that determine the epitaxial registry between the thin film and substrate. We further investigate the microscopic structure of thin films via Raman spectroscopy, demonstrating how the vibrational modes vary as the composition changes and discussing the implications for the crystal structure. We also characterize the samples using electrical transport measurements.

\end{abstract}

\maketitle

\section{Introduction}

\begin{figure}
    \includegraphics[width=8cm]{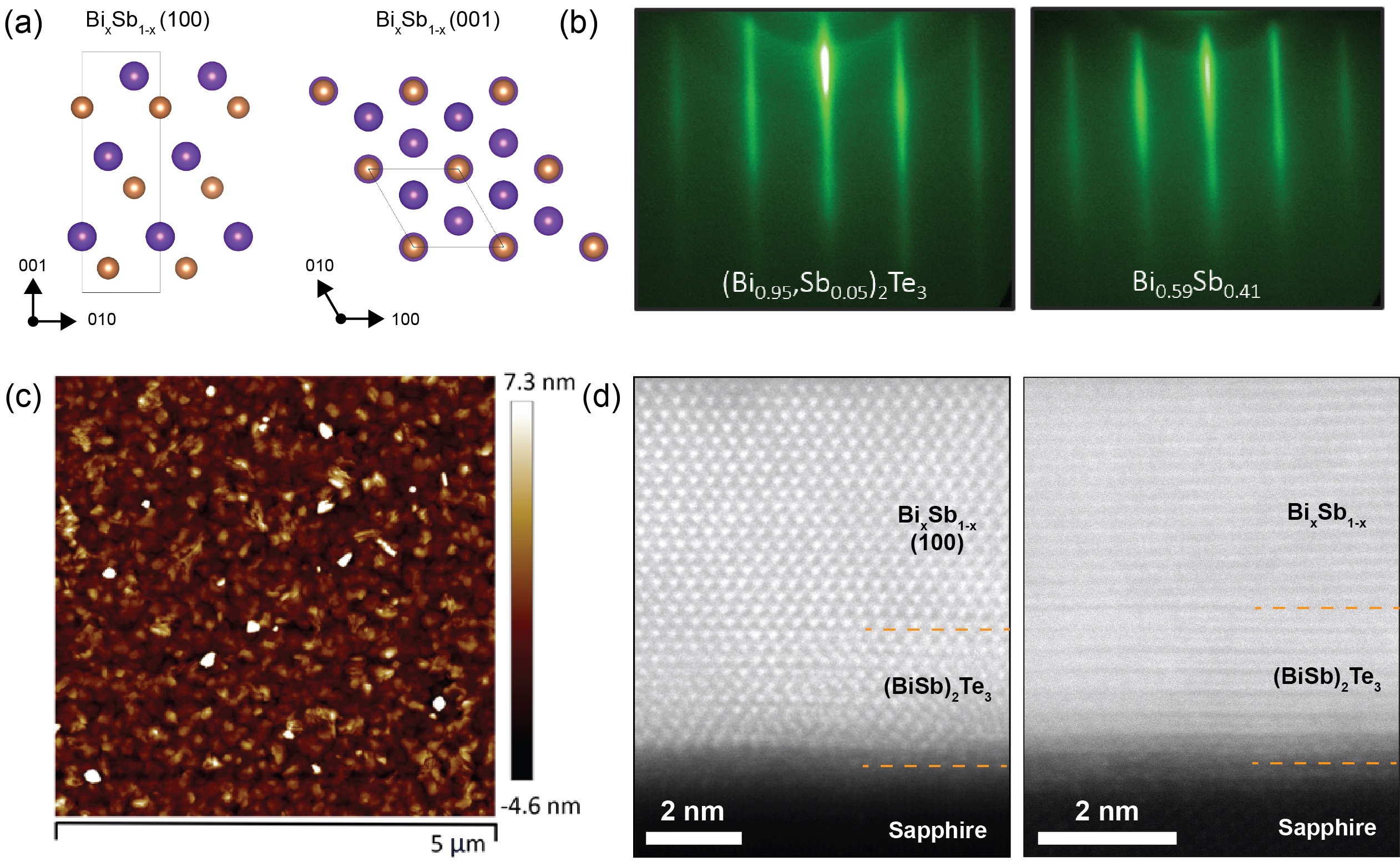}
    \caption{(a) Schematic of the \BiSb~crystal structure along the [100] direction and the [001], growth direction of the film. Violet and dark yellow correspond to Bi and Sb atoms respectively.  (b) RHEED patterns of both the  Bi\textsubscript{0.59}Sb\textsubscript{0.41} layers and the (Bi\textsubscript{0.95},Sb\textsubscript{0.05})\textsubscript{2}Te\textsubscript{3} buffer layer, the electron beam is directed along the $10\bar{1}$ orientation of sapphire. (c) AFM image of the Bi\textsubscript{0.59}Sb\textsubscript{0.41} film over a 5$~\mu$m by 5~$\mu$m region.(d) HAADF-STEM image of Bi\textsubscript{0.84}Sb\textsubscript{0.16} on (Bi\textsubscript{0.95},Sb\textsubscript{0.05})\textsubscript{2}Te\textsubscript{3} buffer layer grown on sapphire substrate with the atomic structure of the film along the [100] directions as seen in the model in (a), and with a $15-20^{\circ}$ in-plane rotation in a different grain. Both show the atomically smooth interface and epitaxial growth of the layers. }
    \label{1}
 \end{figure} 

The \BiSb~alloy first garnered significant attention in the field of semimetals and narrow bandgap semiconductors due to its electric, magnetic, and thermal properties~\cite{mendez1981pressure,brandt1969investigation,brandt1970electron,brandt1971electron,hiruma1980shubnikov,YIM19721141}. While much of this past work focused on the growth and properties of bulk single crystals, the use of molecular beam epitaxy (MBE) had also been explored, demonstrating the growth of \BiSb~thin films on BaF$_2$ and CdTe substrates \cite{morelli_1990,Cho_JVST}. It has long been recognized that the energy bands of this alloy invert as a function of composition leading to an unusual gapless state at the crossing point~\cite{brandt1972investigation,brandt1972investigation2}. This was later understood to lead to topological surface states that were first predicted by Kane and Mele~\cite{kane2005z} and then experimentally realized in bulk single crystals ~\cite{hsieh2008topological,hsieh2009observation}. Recently, there has been renewed interest in
thin films of \BiSb~grown by both MBE ~\cite{yao2019influence,sadek2022structural} and sputtering \cite{fan2020crystal}, motivated by the prospect of exploiting the spin-momentum correlation in the topological Dirac surface states for spin–orbit-torque (SOT) devices. The high electrical conductivity of \BiSb~thin films coupled with a potentially large spin Hall angle makes them attractive for realizing energy efficient SOT memory devices~\cite{ueda2017epitaxial}. Large spin Hall angle $\approx52$ and magnetization switching at ultralow current densities have been reported in \BiSb~films interfaced with the metallic ferromagnet, MnGa \cite{khang2018conductive}, while interfacing with the ferromagnetic semiconductor, (Ga,Mn)As, has shown giant unidirectional spin Hall magnetoresistance~\cite{duy2019giant}. These studies motivate the further exploration of MBE growth of high-quality, single-crystal \BiSb~films on different substrates. As mentioned above, attempts to grow high structural quality \BiSb~thin films on BaF$_2$ and CdTe (111) substrates using MBE date back to the late 1990s \cite{Cho_JVST}.  More recent studies of MBE-grown single crystal \BiSb~thin films have used GaAs (001) as a substrate, yielding \BiSb~(012) growth of either polycrystalline or textured nature, with limited structural characterization details \cite{khang2018conductive,yao2019influence,sadek2022structural}. The renewed interest in \BiSb~thin films for topological spintronics provides a strong motivation for exploring, understanding, and improving the epitaxial growth of single crystal thin films of this material on different substrates.  

In this manuscript, we present a comprehensive study of a series of epitaxial \BiSb~thin films grown by MBE on sapphire (0001) substrates after the deposition of a thin (Bi$_{0.95}$,Sb$_{0.05}$)$_2$Te$_3$ buffer layer. We explore the MBE growth of \BiSb~thin films over the entire composition range, characterizing the microscopic structure, quality, and composition of these films with a comprehensive suite of techniques, including transmission electron microscopy (TEM), x-ray photoelectron spectroscopy (XPS), x-ray diffraction (XRD), atomic force microscopy (AFM), electrical transport, and Raman spectroscopy. The phonon modes in these thin films observed using Raman spectroscopy provide insights into the effect of strain. Our results demonstrate the growth of epitaxial \BiSb~films over the full range of composition;  the controlled synthesis of such films is relevant for the systematic exploration of spintronic devices based upon spin-charge interconversion in this material \cite{Ou_2023}. 

\begin{figure}
    \includegraphics[width=8cm]{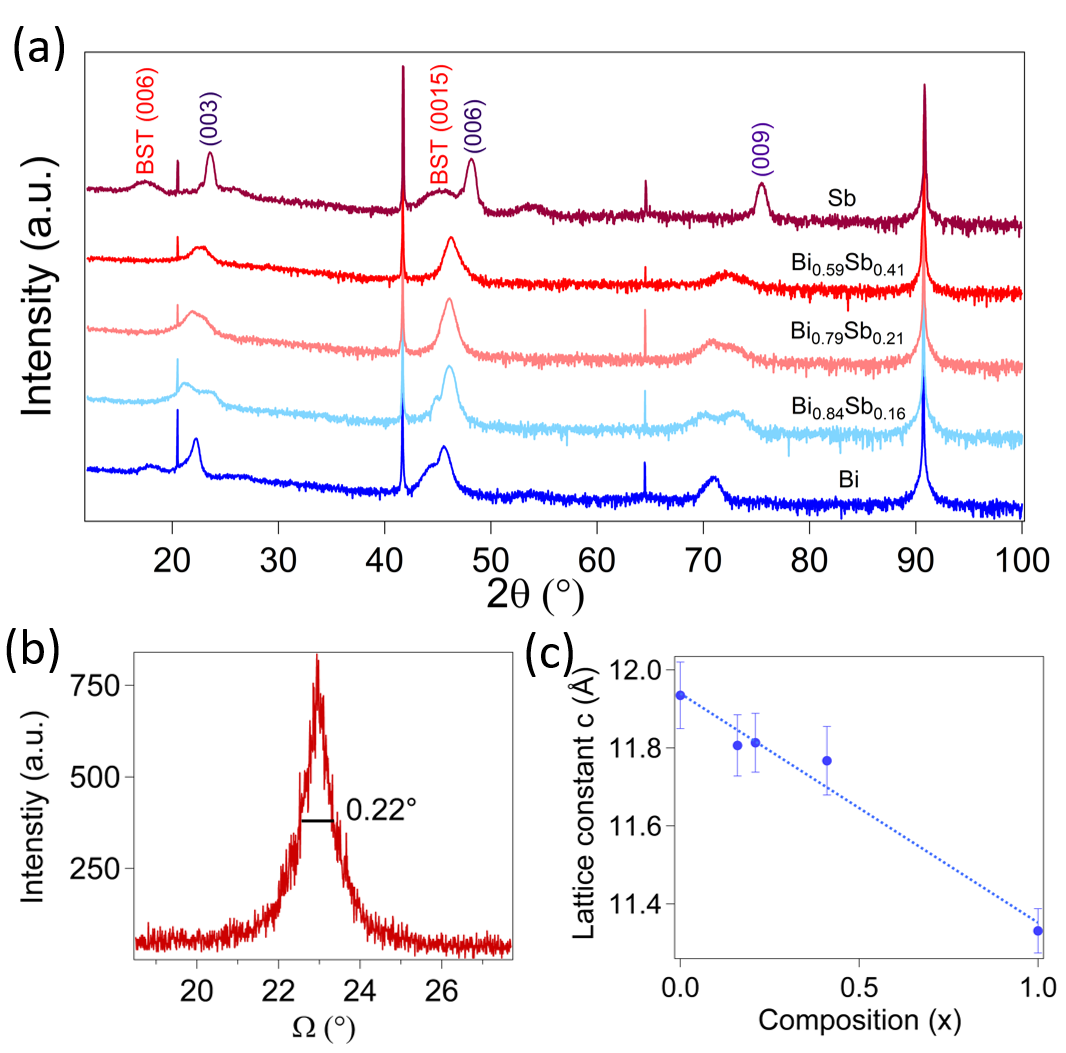}
    \caption{(a) Out-of-plane $2\theta$-$\omega$ XRD measurements of 10 nm-thick \BiSb~films of varying composition grown on sapphire (0001) substrates (samples A-E, as described in Table I). Peaks from the \BiSb~layer, (Bi\textsubscript{0.95},Sb\textsubscript{0.05})\textsubscript{2}Te\textsubscript{3} buffer layer and the substrate layer are detected. (b) A rocking curve of Bi\textsubscript{0.84}Sb\textsubscript{0.16}. (c) Lattice constant $c$ extracted from the $2\theta$-$\omega$ XRD spectrum plotted against composition. The dotted line indicates the trend suggested by Vegard's law.}
    \label{2}
 \end{figure} 
\begin{figure*}
    \includegraphics[width=16cm]{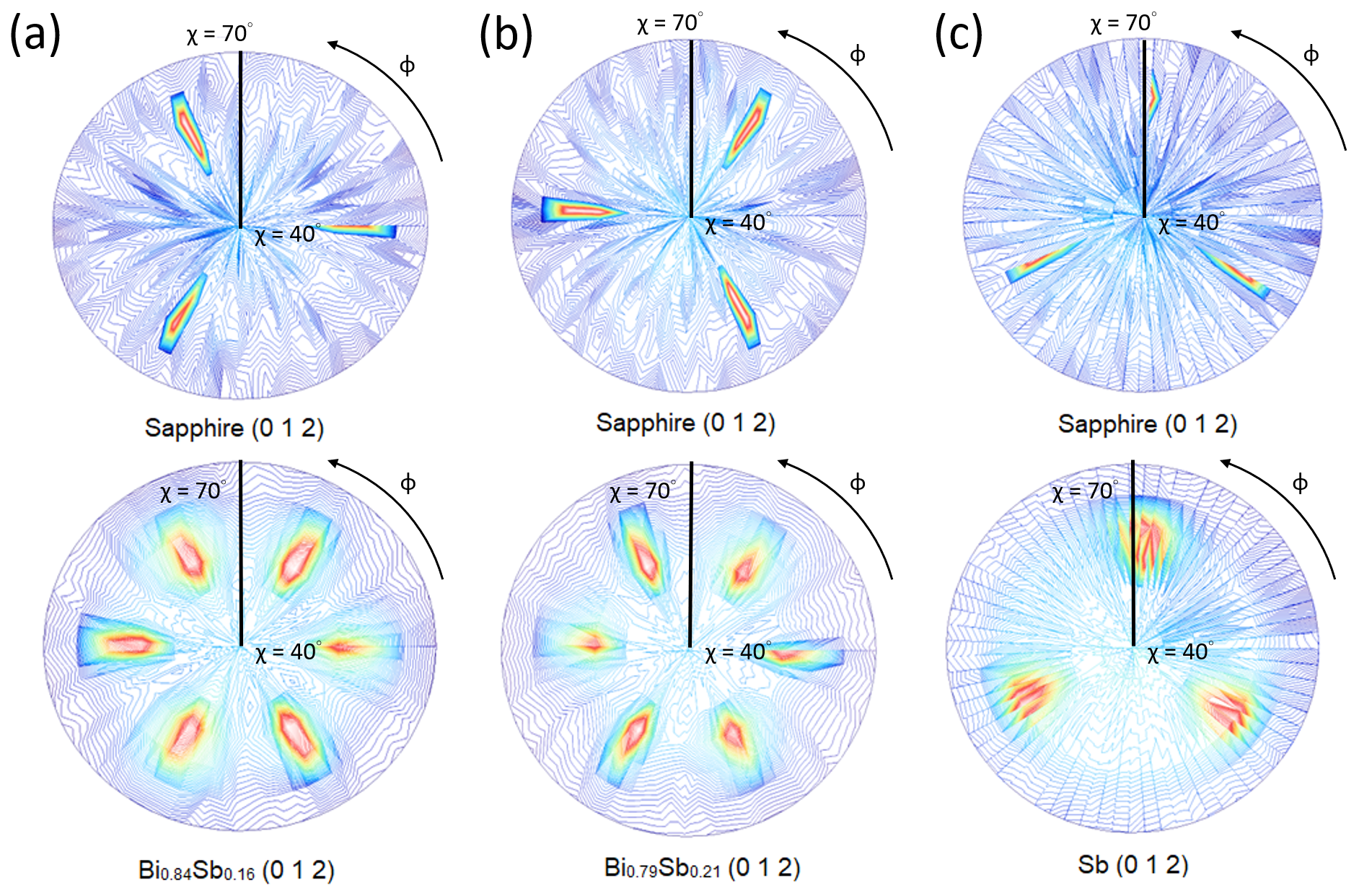}
    \caption{XRD pole figure measurements of the (012) peak in both the substrate and sample layer in (a) Bi\textsubscript{0.86}Sb\textsubscript{0.14}, (b) Bi\textsubscript{0.79}Sb\textsubscript{0.21}, and (c) pure  Sb.}
    \label{3}
 \end{figure*} 

\section{Methods}
We grow our samples on sapphire (0001) substrates using a Scienta Omicron EVO50 MBE system. We first prepare the sapphire substrates by heating up to 1160~$^\circ$C in air for 8 hours to produce an atomically ordered surface, dice it from the back so it can be broken into smaller pieces for future use, and clean it with a series of baths in acetone, isopropanol, DI water, Nanostrip and then DI water again. Once pieces of the substrate are loaded into the MBE system 
(base pressure $<10^{-10}$ mBar), they are outgassed at a substrate heater temperature of 950~$^\circ$C for 1 hour. We then proceed to grow 3 nm of (Bi\textsubscript{0.95},Sb\textsubscript{0.05})\textsubscript{2}Te\textsubscript{3} (in-plane lattice constant $0.437$~nm) as a buffer layer by setting the flux ratio of Bi/Sb to 19 with an overabundance of Te flux. 
The sample temperature is set to 170~$^\circ$C (as measured by an infrared thermal camera, substrate heater temperature $\approx 360~^{\circ}$C) at the start and gradually increased to 200~$^{\circ}$C. Finally, we grow 10 nm of \BiSb~ with different Bi/Sb compositions by adjusting the flux ratio accordingly. A schematic of the crystal structure is shown in Fig. 1(a).  \BiSb~ has a trigonal crystal system and belongs to R$\overline{3}$m space group, with lattice constants $a=b=0.433 - 0.455$~nm (depending on composition), leading to $\approx 3$\% mismatch. The sample temperature is set at 160~$^\circ$C (90~$^\circ$C for Bi) during the growth. The lower growth temperature for pure Bi is chosen based on previous work~\cite{ueda2017epitaxial}, which show epitaxial growth of Bi at lower temperature. We use reflection high energy electron diffraction (RHEED) to monitor the growth. A streaky unreconstructed RHEED pattern is seen during the growth for both the buffer layer and the sample layer (Fig. 1(b)), implying a relatively smooth surface and good epitaxial growth.

\begin{figure}
    \includegraphics[width=8cm]{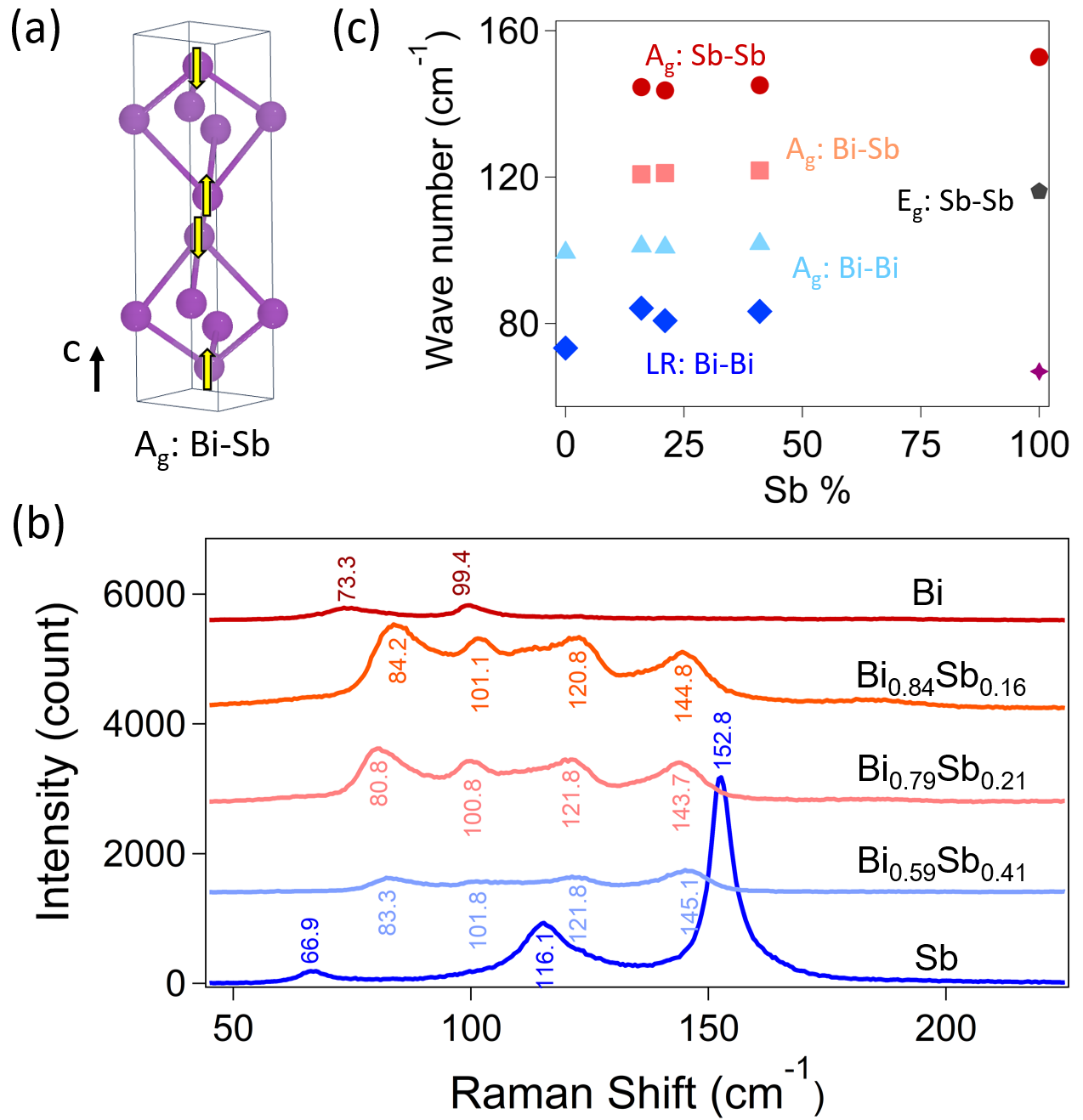}
    \caption{(a) Schematic of the A\textsubscript{g} vibrational mode of Bi-Sb crystal. (b) Raman spectroscopy of \BiSb~films. (c) Peak location of vibrational modes versus Sb composition. LR signified long-range metal like vibration of Bi-Bi.}
    \label{4}
\end{figure} 

  \begin{figure*}
    \includegraphics[width=16cm]{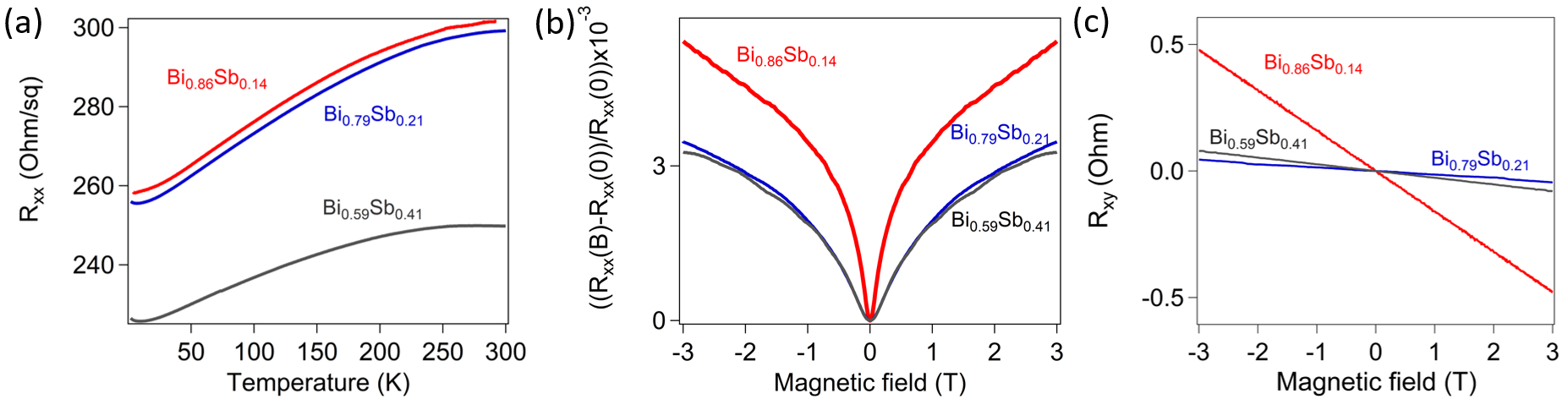}
    \caption{Electrical transport measurements in three \BiSb~films.  Temperature dependence of the sheet resistance for three compositions of Bi and Sb. (b) Normalized longitudinal magnetoresistance, and (c) Transverse magnetoresistance for the same films.  The raw data for $R_{xx}$ and $R_{xy}$ have been symmetrized and antisymmetrized, respectively.}
    \label{5}
 \end{figure*} 
We focus our discussion on five thin films with the same thickness ($\sim$10 nm) and different composition as determined by varying the Bi:Sb flux ratio; the sample composition is verified post-growth using x-ray photoeelectron spectroscopy (XPS) (Table 1). Unlike past work on \BiSb~thin films \cite{morelli_1990,Cho_JVST,sadek2022structural}, we constrain our growth to very thin films since our primary interest is to use these films eventually in SOT devices. 
The thickness and surface topography of these samples are obtained using AFM, as shown in Fig. 1(c). For films of $\sim$10~nm thickness, the RMS surface roughness is $\sim1-2$ nm. We note that this is still somewhat above the requirements for SOT devices, but comparable to our recent work on TaAs thin films, which show giant SOT efficiency~\cite{PhysRevApplied.18.054004}.
We use high angle annular dark-field (HAADF) scanning transmission electron microscopy (STEM) imaging to study the film structure from similarly grown films. Figure 1(d) shows the atomic structure of a \BiSb~ film along the [100] direction from different areas from the film, with the layered structure of the (Bi$_{0.95}$,Sb$_{0.05}$)$_2$Te$_3$ buffer layer visible underneath. Further, the HAADF images show an atomically smooth interface between the \BiSb~ and the buffer layer, confirming a good epitaxial growth between the two.

\begin{table}

\begin{tabular}{|c|c|c|c|c|}
\hline 
Sample & Growth Temperature ($^\circ$C) & Bi/Sb Flux Ratio & Bi\% & Sb\%\tabularnewline
\hline 
\hline 
A & 90 & Pure Bi & 1 & 0\tabularnewline
\hline 
B & 160 & 3.4 & 0.84 & 016\tabularnewline
\hline 
C & 160 & 2.6 & 0.79 & 0.21\tabularnewline
\hline 
D & 160 & 0.97 & 0.59 & 0.41\tabularnewline
\hline 
E & 160 & Pure Sb & 0 & 1\tabularnewline
\hline 
\end{tabular}

\caption{Details of sample growth}
\end{table}

\section{Results}

We first discuss the characterization of our samples using XRD measurements. Figure 2 (a) shows out of plane $2 \theta-\omega$ scans of all the five films. We identify the (003) \BiSb~ peaks at 22$^\circ$, (006) peaks at 45$^\circ$-46$^\circ$, and (009) peaks at 70$^\circ$-75$^\circ$. We can also identify the (Bi\textsubscript{0.95},Sb\textsubscript{0.05})\textsubscript{2}Te\textsubscript{3} (006) peak at around 18$^\circ$ and the (0 0 15) peak at around 52$^\circ$ for pure Bi and Sb films, but these peaks are absent in the alloy samples (samples B,C, and D). The reason for this is not understood. We also confirm good quality growth of our film from the rocking curve, an example of which is shown in Fig. 2(b), showing full-width at half maxima of $0.22^\circ$~ The lattice constant $c$ determined from these XRD scans shows a linear relationship with $x$ (the percentage of Sb) in Fig. 2(c), consistent with Vegard's law~\cite{denton1991vegard}. However, unlike past studies of much thicker films such as \BiSb~films on (111) BaF$_2$ \cite{Cho_JVST}, we find the alloys have a lattice constant which is much closer together compared to their pure counterparts. We will discuss this in more detail later in the manuscript. We further investigate these films using XRD pole figure measurements to understand the epitaxial behavior of the growth. Pole figure measurements can be understood as multiple $\phi$-scans at different $\chi$ angles at a fixed $2 \theta - \omega$ angle. This allows us to focus on non-out-of-plane peaks in different layers of the sample. This provides an understanding of how different layers in the film are aligned with the substrate. Ideally, we want to select reflections in all the layers with the same Miller indices, for example, (012), (104), or  (110). Amongst these, (012) has the strongest signal for the \BiSb~layer. Thus, we  measured the (012) peaks in the substrate, (Bi\textsubscript{0.95},Sb\textsubscript{0.05})\textsubscript{2}Te\textsubscript{3}, and Bi$_{1-x}$Sb$_x$ layers (see Fig. 3). Unsurprisingly, the signal from $3$~nm buffer layer was too weak to be measured. Figure 3 shows that the \BiSb~film and the substrate align on top of each other with a broader $\phi$ spread in the sample layer. Twinning in the sample layer is observed only on the alloy samples, as shown in Figs. 3(a) and (b), with sample peaks 60$^\circ$ apart, and is not observed in the pure Sb sample (Fig. 3(c)). These pole figure scans confirm that the samples are epitaxial and that the pure Bi and Sb films are of higher quality than the alloyed \BiSb~films. 

For further understanding of the lattice dynamics, we investigate the vibrational modes of the films using Raman spectroscopy. Raman spectra of single crystals of Bi$_{1-x}$Sb$_x$ were first investigated by Zitter and Watson~\cite{zitter1974raman}. Lannin further investigated sputtered \BiSb~ thin films in 1979~\cite{lannin1979first}. Both results found five vibrational modes in \BiSb~alloys: $A_g$ and $E_g$ mode of Sb-Sb vibration, $A_g$ mode of Bi-Sb vibration, $A_g$ mode of Bi-Bi vibration, and long-range metallic-like Bi-Bi vibration. The $A_g$ vibrational mode of the Bi-Sb is shown schemtically in Fig. 4(a). The Raman spectra in our films are shown in Fig. 4(b) and the wavenumber of different modes is plotted against composition in Fig. 4(c). We identify the long-range metal-like Bi-Bi vibration mode (blue diamonds) at $73.3$~cm$^{-1}$ in the pure Bi sample and $80-84$~cm$^{-1}$ in \BiSb~ alloy samples; this mode shifts to higher wave number as Sb concentration increases. The $A_g$ mode of the Bi-Bi vibration (blue triangles) is observed around $100$~cm$^{-1}$ for all samples except in the pure Sb sample; this mode remains constant as the Sb fraction changes in the alloy samples (samples B, C, and D). The $A_g$ mode of the Bi-Sb vibration (orange squares) is observed around $120$~cm$^{-1}$ for all \BiSb~ alloy samples; this mode also remains at approximately the same location as the composition is varied in the alloy samples. The $E_g$ mode of the Sb-Sb vibration (black pentagon) is only observed at $116.1$~cm$^{-1}$ in pure Sb; this is most likely due to the lack of enough Sb concentration in the alloy samples. The $A_g$ mode of the Sb-Sb vibration (red circles) is observed at $152.8$~cm$^{-1}$ in the pure Sb sample and at $145$~cm$^{-1}$ in \BiSb~ alloy samples; this shows similar behavior as the $A_g$ mode of the Bi-Bi vibration, and remains approximately at the same position in samples B, C, D, with the mode in the pure Sb film observed at higher wave number. Furthermore, we also identify an extra mode at $66.9$~cm$^{-1}$ (purple stars) in the pure Sb film. This mode does not correspond to any mode previously observed in single crystal \BiSb~ alloys. We believe this corresponds to the $A_g$ mode from Sb$_2$Te$_3$, suggesting the interdiffusion of Te from the (Bi\textsubscript{0.95},Sb\textsubscript{0.05})\textsubscript{2}Te\textsubscript{3} buffer layer. In previous studies, the wave number of the $A_g$ modes had a linear relationship with the composition of the Sb \cite{zitter1974raman,lannin1979first}. In our study, we observe these modes to be located at approximately the same wave number. Our alloy films have a similar lattice constant $c$, as calculated from XRD measurements (Fig. 2(b)). The $A_g$ mode vibrations are along the $c$ direction as shown in the schematic in Fig. 4(a), and thus the wave number of the vibrations also correspond to the lattice constant $c$. This can further be seen in Fig. 2(b), where sample C(79:21) has a slightly larger lattice constant $c$ compared to sample B(84:16) and similarly we also observe a slightly smaller wave number of the $A_g$ in sample C compared to sample B. We believe this is because presence of strain in our sample~\cite{ni2008uniaxial}.

Finally, we investigate electrical transport in \BiSb~thin films of varying composition grown under similar conditions. We carry out these measurements using mechanically scratched Hall bars with channel dimensions 1 mm $\times 500 ~\mu$m. The films show similar qualitative and quantitative behavior. The sheet resistance ($R_{xx}$) vs. temperature ($T$) in all samples shows metallic behavior (Figs. 5(a)). Longitudinal magnetoresistance measurements at $T=4.2$~K shows a positive magnetoresistance whose field-dependence is consistent with that expected from weak antilocalization (Figs. 5(b). The carrier density ($n$) for all the samples is extracted using Hall measurements (Figs. 5(c)). The samples are electron-doped with $n\approx 10^{20}-10^{22}$~cm$^{-3}$. The mobility is calculated using $\mu=\frac{\sigma}{ne}$, where $\sigma$ is the conductivity and $e$ is the electronic charge. The Fermi wavelength ($k_f$), calculated using  $k_f=(3\pi^2n)^{\frac{1}{3}}$ is $\approx 10^{-9}$~m$^{-1}$ while the mean free path ($l_m$), calculated using $l_m=\mu \hbar k_f/e$ varies between $10^{-8}-10^{-9}$~m. $k_fl_m$ varies between $10-25$ in these films.  The parameters calculated from transport measurements are provided in Table II. Note that the relatively low Hall mobilities extracted here can be attributed to the thin nature of our films, where interfacial defects which act as trapping-detrapping centers~\cite{islam_adma.202109671} and grain boundaries from twinning as observed in XRD (Fig. 3), can lead to enhanced scattering compared to thicker films: for example, the MBE growth of \BiSb~thin films of about 60 nm thickness on GaAs substrates also shows Hall mobilities of order 100 cm$^2$/(V.s) \cite{sadek2022structural}. 

\begin{table}
\begin{tabular}{|c|c|c|c|c|c|}
\hline 
Sample & $n$~(cm$^{-3}$) & $\mu$~(cm$^2$/Vs) & $k_f$~($10^9$~m$^{-1}$) & $l_m$~(m) & $k_f l_m$ \tabularnewline
\hline 
\hline 
A & $-2.48\times10^{21}$ & 1 & 9 & $5.2\times10^{-10}$ & $4.7$\tabularnewline
\hline 
B & $-1.9\times10^{21}$ & $12.7$ & $3.8$ & $3.2\times10^{-9}$ & $12.3$\tabularnewline
\hline 
C & $-5.7\times10^{20}$ & $67.7$ & $2.4$ & $1.07\times10^{-9}$ & $25.8$\tabularnewline
\hline 
D & $-1.28\times10^{22}$ & $2.2$ & $7.18$ & $1.04\times10^{-9}$ & $7.5$\tabularnewline
\hline 
E  & $-6.7\times10^{20}$ & $51.4$ & $2.7$ & $9.1\times10^{-9}$ & $24.8$\tabularnewline
\hline 
\end{tabular}

\caption{Details of transport measurements}
\end{table}
\section{Conclusion}
In conclusion, we used MBE to demonstrate the epitaxial growth of a series of \BiSb~thin films over the entire alloy composition range on sapphire (0001) substrates with a thin (Bi\textsubscript{0.95},Sb\textsubscript{0.05})\textsubscript{2}Te\textsubscript{3} buffer layer. Detailed structural characterization of the films via XRD and HR-TEM confirm high-quality epitaxial registry with the substrate, while the composition of these films was verified using XPS. We also probed the effect of strain (and possible interdiffusion) using Raman spectroscopy and probed their electrical properties using low temperature transport. Our growth protocol establishes sapphire (0001) as an attractive substrate for MBE growth of \BiSb~and provides a route for systematic studies of the spin-charge conversion and SOT  devices by interfacing such films with a ferromagnetic overlayer \cite{Ou_2023}.

\begin{acknowledgments}
This project was supported by SMART, one of seven centers of nCORE, a Semiconductor Research Corporation program, sponsored by the National Institute of Standards and Technology (NIST) (YS, YO, NS, SG, KAM), NSF Grant No. DMR-2309431 (SG, KAM), and by the Penn State Two-Dimensional Crystal Consortium-Materials Innovation Platform (2DCC-MIP) under NSF Grant No. DMR-2039351 (YO, SI, AR, NS). Parts of this work were carried out in the Characterization Facility, University of Minnesota, which receives partial support from the NSF through the MRSEC (Award Number DMR-2011401) and the NNCI (Award Number ECCS-2025124) programs (SG, KAM).
\end{acknowledgments}


\providecommand{\noopsort}[1]{}\providecommand{\singleletter}[1]{#1}%

\end{document}